\documentclass{aa}  

\usepackage{graphicx}
\usepackage{arydshln}
\usepackage{txfonts}
\usepackage{xcolor}
\usepackage[normalem]{ulem}
\begin{document}

   \title{Formation of bi-lobed shapes by sub-catastrophic collisions}

   \subtitle{A late origin of comet 67P/C-G's structure}

   \author{M. Jutzi, W. Benz                    }

   \institute{Physics Institute, University of Bern, NCCR PlanetS,
              Sidlerstrasse 5, 3012 Bern, Switzerland\\
              \email{martin.jutzi@space.unibe.ch; willy.benz@space.unibe.ch}
                           }

   \date{Received -- ; accepted --}

 
  \abstract
   {The origin of the particular shape of a small body like comet 67P/Churyumov-Gerasimenko (hereafter 67P/C-G) is a topic of active research. How and when it acquired its peculiar characteristics has distinct implications on the origin of the solar system and its dynamics}
   {We investigate how shapes like the one of comet 67P/C-G can result from a new type of low-energy, sub-catastrophic impacts involving elongated, rotating bodies. We focus on parameters potentially leading to bi-lobed structures. We also estimate the probability for such structures to survive subsequent impacts.} 
   {We use a smooth particle hydrodynamics (hereafter SPH) shock physics code to model the impacts, the subsequent reaccumulation of material and the reconfiguration into a stable final shape. The energy increase as well as the degree of compaction of the resulting bodies are tracked in the simulations.}
   {Our modelling results suggest that the formation of bi-lobed structures like 67P/C-G is a natural outcome of the low energy, sub-catastrophic collisions considered here.}
   {Sub-catastrophic impacts have the potential to alter the shape of a small body significantly, without leading to major heating or compaction. The currently observed shapes of cometary nuclei, such as 67P/C-G, maybe a result of such a last major shape forming impact.}


   \maketitle
%

\section{Introduction}\label{sec:introduction}
Whether cometary nuclei structures as observed today are pristine and preserve a record of their original accumulation, or are a result of later collisional or other evolutionary processes is still much debated \citep[e.g. ][]{Weissmann:2004wa,Mumma:1993mw,Sierks:2015sb,Rickman:2015wu,Morbidelli:2015vm,Davidsson:2016ds}.

Based on data from the European Space Agency's Rosetta rendezvous mission \citep{Sierks:2015sb}, it was suggested that the particular bi-lobe structure of comet 67P/C-G was formed during the early stages of the Solar System \citep{Massironi:2015ma}, possibly by low velocity accretionary collisions \citep{Jutzi:2015ja} and therefore should be considered as a primordial body. On the other hand, \citet{Morbidelli:2015vm} show that in the "standard scenario" of the early dynamical evolution of the Solar system, an object of the size of comet 67P/C-G would have experienced a high number of catastrophic collisions and thus could not have survived. This study has been improved and a detailed analysis of the survival probability of a 67P/C-G-like object is presented in a companion paper (Jutzi et al., 2016, submitted; hereafter Paper I). It is found that even in the scenario without a long-lasting primordial disc, comet 67P/C-G could not have conserved its primordial shape and a large number of shape-changing collisions should have occurred during its lifetime. This conclusion holds also true for generic bi-lobe structures, which might evolve further through a fission-merging cycle as recently suggested by \citet{Scheeres:2016sh}.  These results strongly indicate that neither the current shape of  comet 67P/C-G nor its two individual lobes can actually be primordial. 
 
Alternatively, 67P/C-G-like bi-lobe structures could be the result of collisional disruptions of somewhat larger bodies taking place at a later stage in the history of the solar system \citep[e.g. ][Paper I]{Rickman:2015wu,Morbidelli:2015vm,Marchi:2015mr}. During these later stages, typical relative velocities between small bodies have grown much larger than their mutual escape velocity ($V>>V_{esc} \approx 1$ m/s) of kilometer-sized bodies and therefore direct bi-lobe formation by collisional mergers of similar sized bodies \citep{Jutzi:2015ja} is no longer a viable mechanism. However, suitable low relative velocities could be found again between fragments/aggregates following a higher-velocity collisional disruption of a larger parent body. In this secondary bi-lobe formation scenario, the dispersed material following a catastrophic collision re-accumulates in two gravitationally bound bodies which subsequently collide at low speed. Re-accumulation has been extensively studied for asteroids and results show that such behaviour is indeed occurring \citep[e.g.][]{Michel:2001mb,Michel:2003mb,Michel:2013mr}. A study of large-scale disruption and sub-sequent reaccumulation in the case of cometary parent bodies is currently in progress (Schwartz et al., 2016, in prep.). 

However, if the currently observed 67P/C-G structure did form as a result of a late catastrophic disruption of a larger parent body, the resulting shape must then survive until today. In other words, from the time of formation until present day it cannot have experienced a single subsequent shape-changing collision. In this context, the difference between the specific impact energy $Q$ required to catastrophically disrupt the parent body of 67P/C-G and the energy sufficient to change its shape $Q_{reshape}$ is a crucial quantity, as it determines the relative number of such events. On average, the larger the specific energy differences, the larger the number of shape changing impacts compared to the number of disruption events. Catastrophic disruptions are usually characterised by the specific impact energy $Q^*_D$, which leads to the escape of half of the initial mass involved in the collision.  As it is shown in Paper I, the ratio $Q^*_D$/$Q_{reshape}$ is very large, of the the order of $10^3$. Hence, there are on average many more shape changing collisions than disruptive collisions and it is unlikely that 67P/C-G could have conserved its shape, unless the disruption of the parent body has occurred very recently.   

As we will show in this paper,  67P/C-G-like bi-lobe structures could also emerge from low energy, sub-catastrophic impacts on parent bodies of cometary size. It turns out that the specific energy $Q_{sub}$ for these types of impacts is much closer to the specific re-shape energy $Q_{reshape}$ and hence the collision frequency difference between the two events is much smaller. This translates directly in a larger survival rate and therefore a higher probability to observe a shape like 67P/C-G. 

As suggested recently \citep{Scheeres:2016sh}, comets with two-component shapes might enter a fission-merging cycle, once they enter the inner solar system and experience changes in their spin rate.  Hence, its is possible that bi-lobe structures undergo episodic shape changes and the present day observed shape is simply the result of the last of such episodes. Even though this scenario adds the possibility for a time dependent macroscopic shape change, it is important to realise that a two-component object of cometary size must exist to begin with. Hence, even in this case the question of the long-enough survival of bi-lobe shapes is a central question (Paper I). 

Observational results, such as the abundance of supervolatiles (CO, CO$_2$, N$_2$) \citep[e.g.][]{Haessig:2015he,LeRoy:2015ra}, the detection of primordial molecules \citep{Bieler:2015ba}, and the evidence for a low formation temperature \citep{Rubin:2015ra} suggest that comet 67P/C-G cannot have experienced any substantial, global scale heating after its formation. Further constraints include the high porosity and the observed homogeneity of the nucleus, which appears to be constant in density on a global scale without large voids  \citep{Paetzold:2016pa}. As a consequence, any proposed formation mechanism of bi-lobe shapes must be able to operate within very tight constraints on energy input and compaction of porous material. 

In this paper we investigate the final shapes resulting from a new type of low-energy, sub-catastrophic impacts on elongated, rotating bodies that meets the constraints mentioned above. We carry out a set of 3D smooth particle hydrodynamics simulations of impacts to investigate the possibility to forming bi-lobe structures reminiscent of those objected for cometary nuclei. In section \ref{sec:model} we present our model approach and describe the setup and initial conditions. Representative bi-lobe forming collision are presented in section \ref{sec:examples}. The results (final shapes) of our parameter space exploration are shown in section \ref{sec:shapes}; heating and compaction effects are discussed in section \ref{sec:effects}. In section \ref{sec:probability} we investigate the probabilities for a 67P/C-G-like structures formed in such a manner to avoid destruction by subsequent shape changing collisions.

Finally we discuss our bi-lobe formation model in the context of the question of how primordial comets are (section \ref{conclusions}).  

\section{Bi-lobe formation by sub-catastrophic impacts}\label{sec:model}
\subsection{Motivation and assumptions}
Remote observations of cometary nuclei suggest that a large fraction of these objects have elongated rather than sphere-like shapes \citep{Lamy:2004lt}. This is interesting as it turns out that elongated bodies are naturally easier to 'split' in two components than spherically symmetric bodies. This is even more true when they are rotating around their short axis, and centrifugal forces act in opposite directions at each end of the body. Impacts on such elongated rotation bodies might therefore act as a splitting mechanism leading first to two distinct bodies which can potentially form a binary system or eventually merge together forming a bi-lobed body.

To study under which conditions such splitting can indeed produce such bi-lobed structures, we investigate the effects of impacts on rotating ellipsoids. We use axis ratios and rotation periods which are consistent with the observed range of values typical for cometary nuclei \citep{Lamy:2004lt}. We consider impacts in the sub-catastrophic regime for which most of the mass remains bound and accumulates on the main body (possibly made of two components). The impact scenario investigated here is very different from the case of a catastrophic break-up of a large parent body (Schwartz et al., 2016, in prep.) where the largest re-accumulated remnants contain only a small fraction of the initial mass. However, such remnants of catastrophic disruptions might have properties (elongated and rotating) similar to the targets considered here.

\subsection{Modelling approach}
The modelling approach used here is the same as in Paper I. We use a parallel smooth particle hydrodynamics (SPH) impact code \citep{Benz:1995hx,Nyffeler:2004tz,Jutzi:2008kp,Jutzi:2015gb} which includes self-gravity as well as material strength models. To avoid numerical rotational instabilities, the scheme suggested by \citet{Speith:2006} has been implemented.

In our modelling, we include an initial cohesion $Y_0$ $>$ 0 and use a tensile fracture model \citep{Benz:1995hx}, using parameters that lead to an average tensile strength $Y_T$ $\sim$ 100 Pa. To model fractured, granular material, a pressure dependent shear strength (friction) is included by using a standard Drucker-Prager yield criterion \citep{Jutzi:2015gb}. As shown in \citet{Jutzi:2015gb,Jutzi:2015ux}, granular flow problems are well reproduced using this method. 
Porosity is modelled using a P-alpha model \citep{,Jutzi:2008kp} with a simple quadratic crush curve defined by the parameters $P_e$, $P_s$, $\rho_{0}$, $\rho_{s0}$ and $\alpha_{0}$. Following the approach in Paper I, we also introduce the density of the compacted material as $\rho_{compact}$ = 1980 kg/m$^3$ to define the initial distention  $\alpha_{0}=\rho_{compact}/\rho_0=4.5$ corresponding to an initial porosity of $1-1/\alpha_{0} \sim 78\%$.
We apply a modified Tillotson equation of state (EOS; e.g. \citealp{Melosh:1989}) with parameters appropriate for water ice. As in Paper I, we use a reduced bulk modulus, corresponding to an elastic wave speed of $c_e$$\sim$ 0.1 km/s. 

The relevant material parameters used in the simulations are provided in Table \ref{table:matparam}. 

\begin{table*}
\caption{Material parameters used in the simulations. Crush curve parameters $P_e$ and $P_s$  \citep{,Jutzi:2008kp}, density of matrix material $\rho_{s0}$, initial bulk density $\rho_{0}$, density of the compacted material $\rho_{compact}$, initial distention $\alpha_{0}$, bulk modulus $A$, friction coefficient $\mu$, cohesion $Y_0$, average tensile strength $Y_T$. }           
\label{table:matparam}      
\centering                        
\begin{tabular}{l c c c c c c c c c c}        
\hline\hline            
 $P_e$ (Pa) & $P_s$ (Pa) & $\rho_{s0}$ (kg/m3) & $\rho_{0}$ (kg/m3) & $\rho_{compact}$ (kg/m3)& $\alpha_{0}$ & $A$ (Pa)& $\mu$	 & $Y_0$ (Pa) & $Y_T$ (Pa) \\    
\hline                        
 $10^3$ & $10^5$ & 910 & 440 & 1980 & 4.5 & 2.67$\times10^6$ & 0.55 & $10^3$ & $10^2$ \\
   \hline                               
\end{tabular}
\end{table*}

\subsection{Setup and initial conditions}\label{sec:setup}

\begin{figure}
   \centering
   \includegraphics[width=\hsize]{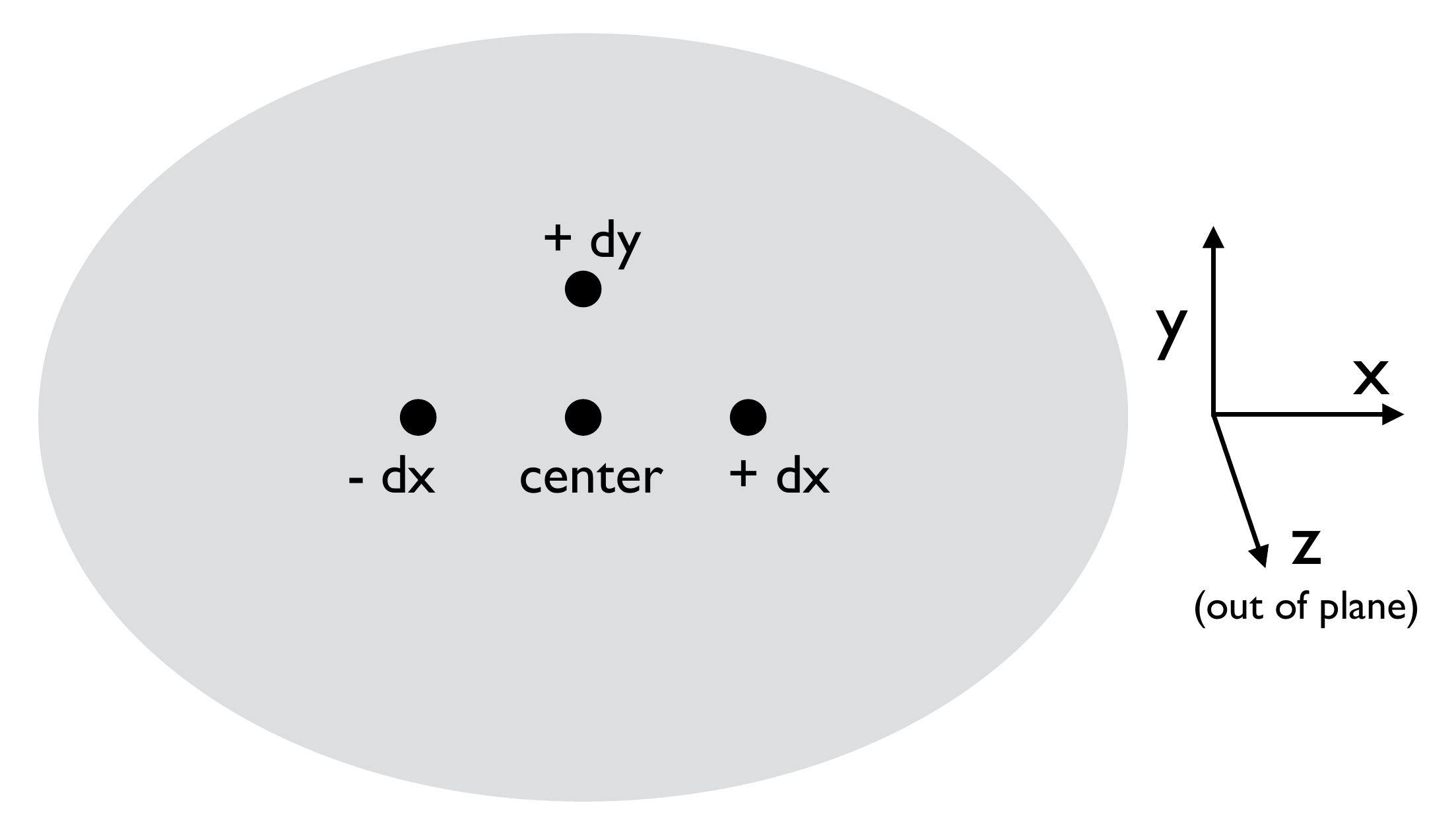}
   \caption{Impact geometries. We use an ellipsoid with a constant length of $L$ = 5.04 km and two different axis ratios (0.4 and 0.7). The ellipsoids are rotating with a rotation axis either along the $y$ or $z$ direction. Various impact points are investigated, as illustrated in the plot. The distance $|dx|=|dy| = $ 670 m.}
     \label{fig:impgeo}
    \end{figure}

There is an infinite number of combinations of impact parameters in the case of non-spherical, rotating targets. Therefore, for practical reasons, we have to limit the size of the initial parameter space that can be investigated. The aim of this paper being more a demonstration of principle than a complete investigation of all possibilities,  we limit ourselves to a few promising cases of relatively central collisions. However, a larger sample of various rotation rates, orientations as well as a range of impact locations have been investigated. The parameters defining the impact  geometries used in this study are illustrated in Figure \ref{fig:impgeo}. 

Target and impactor have both the same initial material properties including their initial bulk density which is set to $\rho_0\sim$ 440 kg/m$^3$.

We use two different ellipsoidal targets both of length $L=5.04$ km but differing from each other by their axis ratios  (0.4 and 0.7).  With the assumed initial density, the two ellipsoids have an initial masses of 4.7$\times$$10^{12}$ kg and 1.45$\times$$10^{13}$ kg, respectively. The impactor size is $R_p$ = 100 m (in collisions involving the target with an axis ratio of 0.4) and $R_p$ = 200 m (in collisions involving the target with an axis ratio of 0.7). The corresponding projectile masses are $M_p$ = 1.8$\times$$10^9$ kg and  $M_p$ = 1.4 $\times$$10^{10}$ kg, respectively. The impact velocities are chosen in the range of 200-300 m/s. We note that the sizes of the impactors as well as the collision velocities are motivated mainly by numerical rather than physical reasons. Smaller impactors at higher speeds would not be well resolved (spatially) in our simulations and would also require smaller timesteps to avoid unphysical oscillations. Even at a relatively modest resolution ($\sim 10^5$ SPH particles), these simulations are quite challenging even for a parallelised code. This is because the simulations have to extend over a very large number of dynamical timescales (and hence involve a very large number of timesteps) before the final structure of the resulting object can be determined (typically one day real time). We note, however, that the impact velocities considered here are super-sonic (sound speed $\sim$ 100 m/s) and the projectile small enough for the impact to be taking place in the so-called point source regime. Hence, our results can be scaled to larger impact speeds using appropriate scaling laws. 

 \begin{figure*}
   \centering
   \includegraphics[width=0.99\hsize]{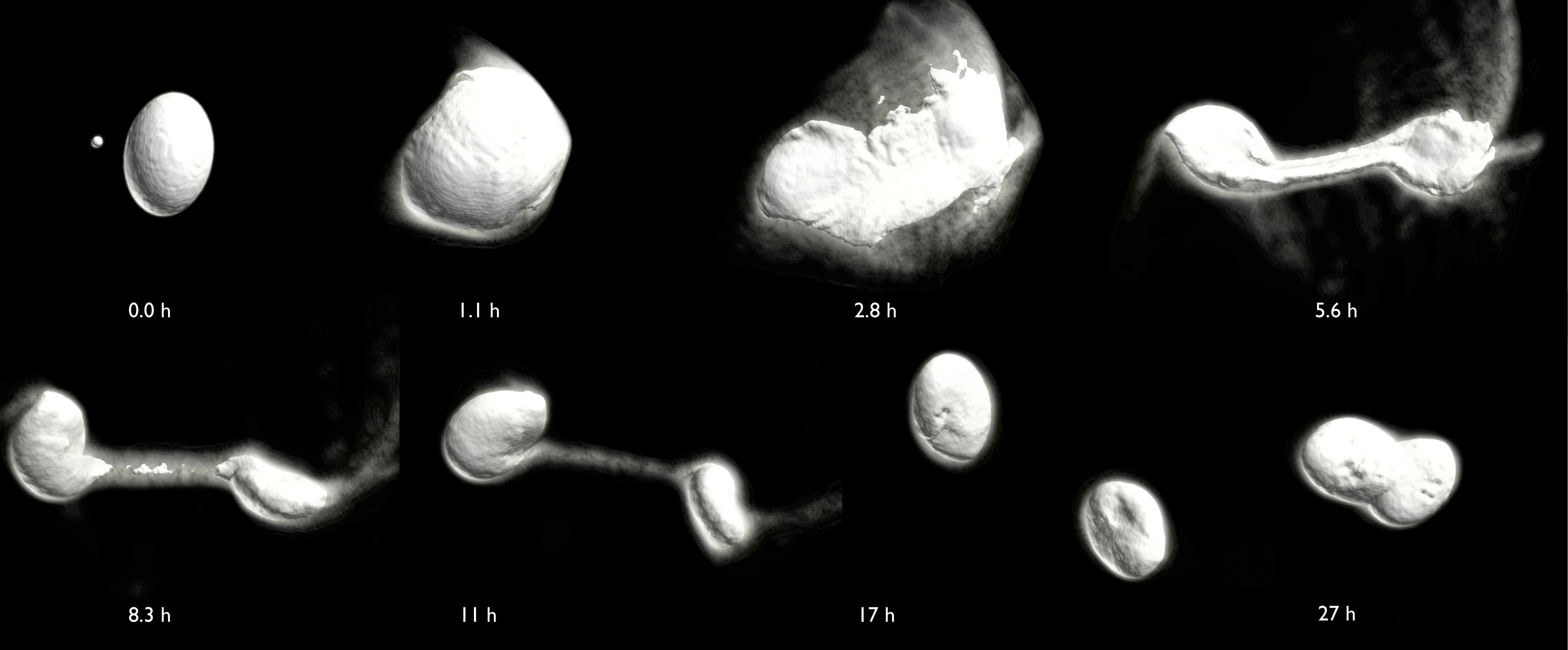}
   \caption{Comet 67P/C-G shape formation by sub-catastrophic collisions. Shown is an example of an SPH calculation of an impact on a rotating ellipsoid. After the initial disruption, subsequent re-accumulation leads to the formation of two lobes. This processes may include the possible formation of layers. The two lobes are gravitationally bound and collide with each other within $\sim$ one day forming a bi-lobed structure. The final shape is also shown in Figure \ref{fig:shapes0.7}. Initial conditions: off axis (impact point + $dy$; see Figure \ref{fig:impgeo}) impact of a $R_p$ = 200 m impactor with a velocity of $V$ = 300 m/s on a target with axis ratio 0.7 (mass $M_t$ = 1.4$\times$$10^{13}$ kg) and rotation period of $T$ = 6 h. The rotation axis is along the $y$-axis (see Figure \ref{fig:impgeo}).}
     \label{fig:snapshots0.7}
    \end{figure*}
    
     \begin{figure*}
   \centering
    \includegraphics[width=0.99\hsize]{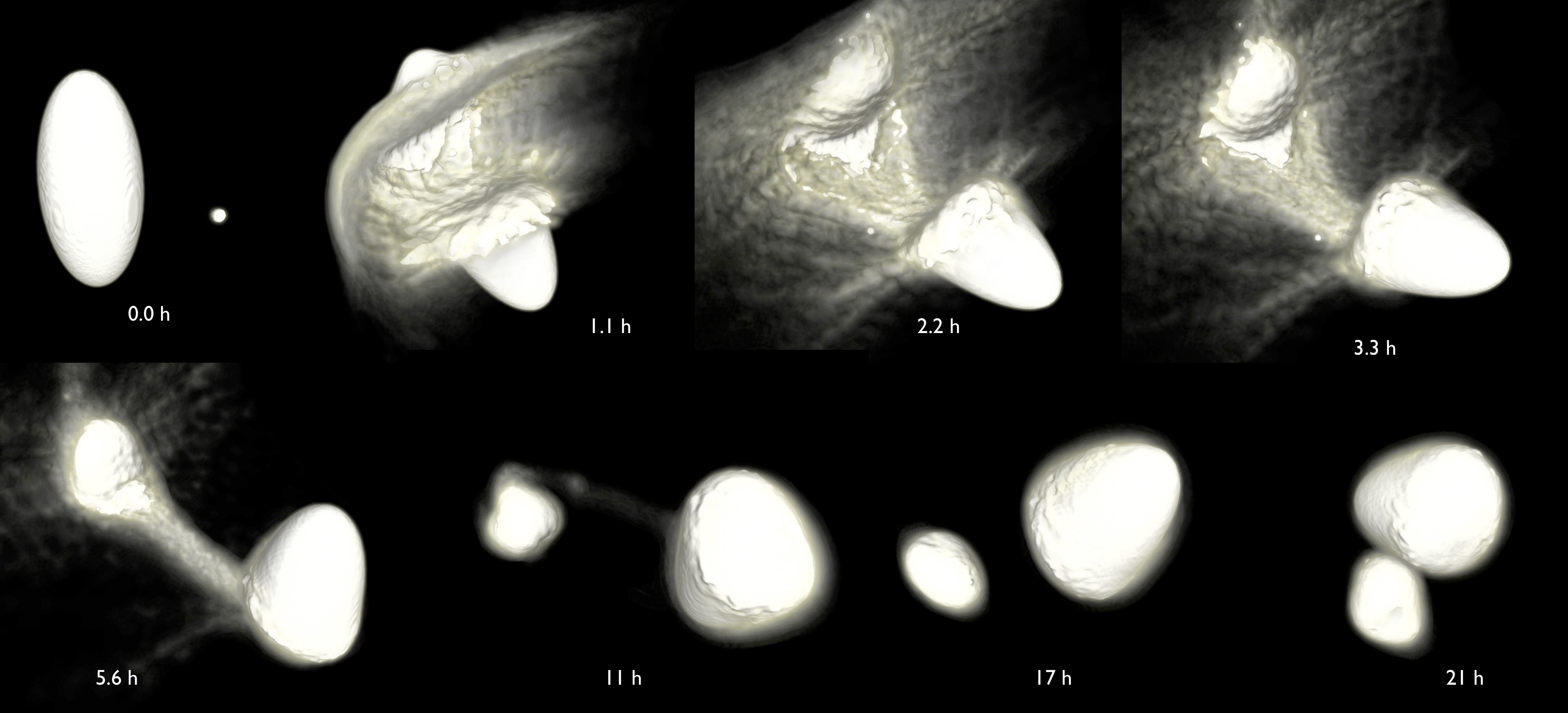}
   \caption{Same as Figure \ref{fig:snapshots0.7} but showing a case with a more elongated target, different rotation axis and impactor properties. The final shape is shown in Figure \ref{fig:shapes0.4}. Initial conditions: off axis (impact point + $dx$; see Figure \ref{fig:impgeo}) impact of a $R_p$ = 100 m impactor with a velocity of $V$ = 250 m/s on a target with axis ratio 0.4 (mass $M_t$ = 4.7$\times$$10^{12}$ kg) and rotation period of $T$ = 9 h. The rotation axis is is along the -$y$ axis (see Figure \ref{fig:impgeo}).}
     \label{fig:snapshots0.4}
    \end{figure*}

\begin{figure*}
   \centering
   \includegraphics[width=0.9\hsize]{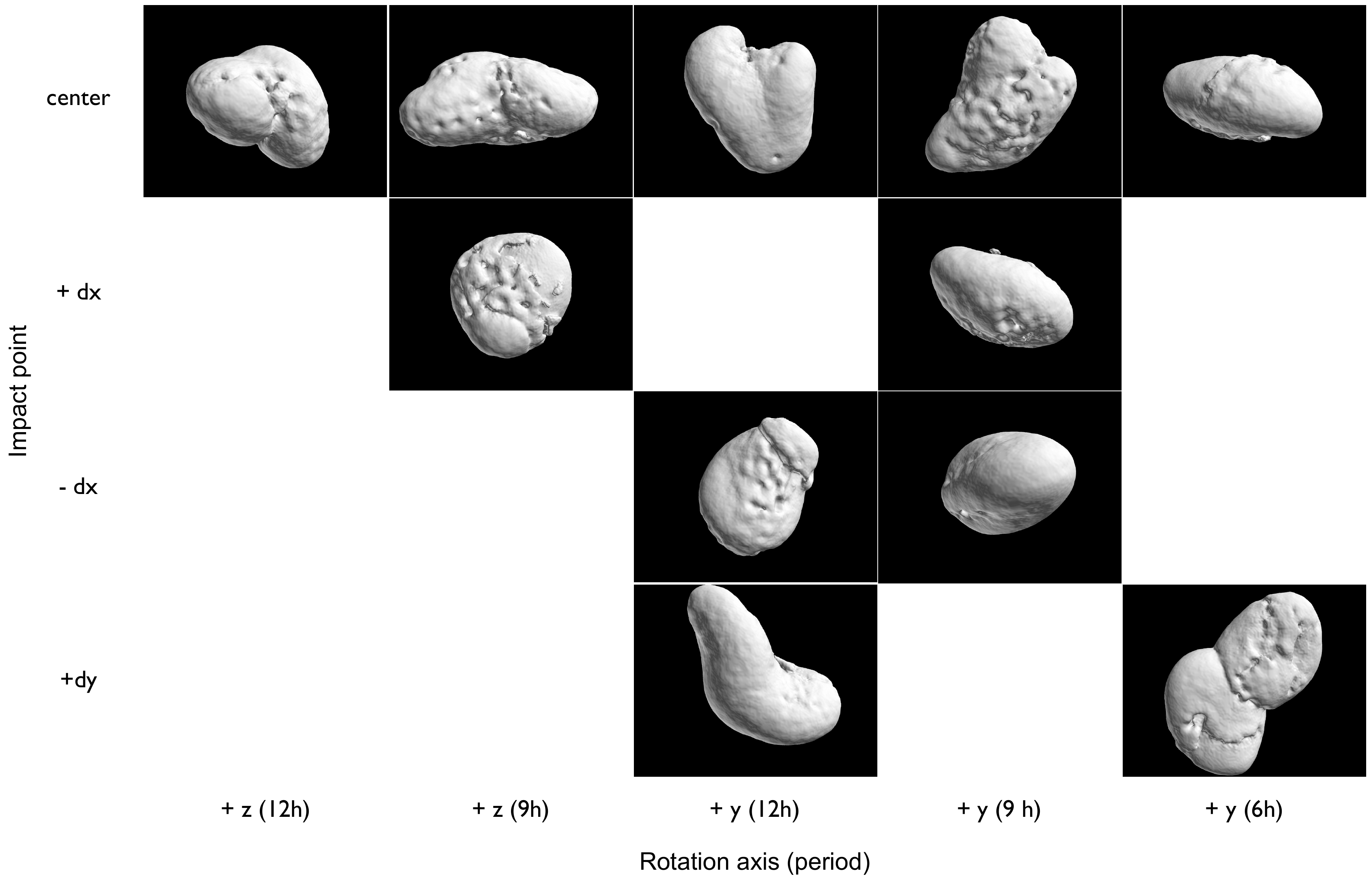}
   \caption{Shapes resulting from sub-catastrophic disruptions of rotation ellipsoids. Shown are the results for different impact positions, rotation axis and periods (see Figure \ref{fig:impgeo} for the impact geometries). The impact velocity is $V$ = 300 m/s and the impactor size  $R_p$ = 200 m. The initial target mass is 1.45$\times$$10^{13}$ kg and the target axis ratio is 0.7. The mass of the final bodies is of the order of $\sim$ 70-80 \% of the initial mass. The case shown in Figure \ref{fig:snapshots0.7} has the coordinates [+$y$(6h),+$dy$].}
     \label{fig:shapes0.7}
    \end{figure*}
    
\begin{figure*}
   \centering
   \includegraphics[width=0.9\hsize]{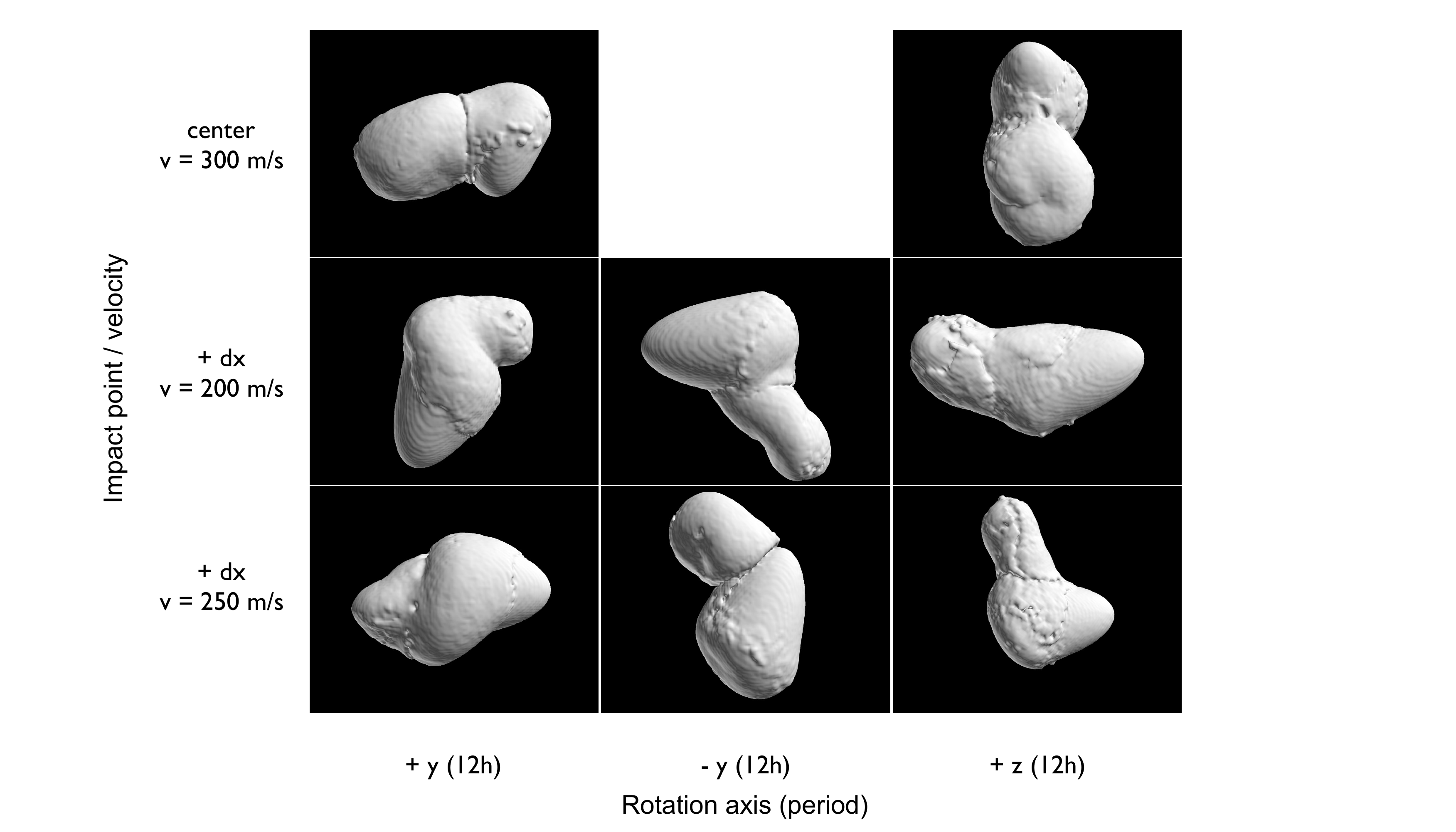}
   \caption{Shapes resulting from sub-catastrophic disruptions of rotation ellipsoids. Shown are the results for different impact positions, velocities and rotation periods and axis (see Figure \ref{fig:impgeo}). The impact velocity is labeled in the plot; the impactor size is  $R_p$ = 100 m. The initial target mass is 4.7$\times$$10^{12}$ kg and the target axis ratio is 0.4. The mass of the final bodies is of the order of $\sim$ 80-90 \% of the initial mass. The case shown in Figure \ref{fig:snapshots0.4} has the coordinates [-$y$(12h),+$dx$ with $v$=250/ms].}
     \label{fig:shapes0.4}
    \end{figure*}
    
\section{Results}
\subsection{Two examples of bi-lobe forming collisions}\label{sec:examples}
We present snapshots from the simulation of two representative cases of bi-lobe forming collisions in Figures  \ref{fig:snapshots0.7} and \ref{fig:snapshots0.4}.  Due to the elongated nature of the targets, the immediate post-impact mass distribution is concentrated at two locations. As a result, subsequent re-accumulation due to gravity leads to the formation of two distinct bodies. Centrifugal forces due to target rotation enhance the initial separation of these two masses. Material re-accumulating at low relative velocity during this process might also lead to the formation of layered structures on each individual body, such as observed on 67P/C-G  \citep{Massironi:2015ma} and computed by \citet{Jutzi:2015ja}. Finally, the two main gravitationally bound bodies eventually collide with each other at low velocity (determined by their mutual gravity and hence of order $\sim 1$ m/s) within $\sim$ one day forming a 67P/C-G-like bi-lobe shape.

We note that the formation of a bi-lobe structure in such low velocity collisions ($V \sim V_{esc}$) of two gravitationally bound objects is consistent with previous results \citep{Jutzi:2015ja}. While these studies considered these low velocity impacts to occur during the early days of the solar system before small bodies became scattered by growing planets, the impact scenario presented here provides additional possibilities for such low velocity collisions to occur much later in the history of the solar system.  

Due to numerical limitations (section \ref{sec:setup}), the velocities considered are at the lower limit of the expected average velocities in the initial left-over planetesimal disc and are significantly lower than average velocity after disc dispersal \citep{Morbidelli:2015vm}. However, the critical energies $Q_{sub}$ for sub-catastrophic bi-lobe formation for different impact velocities (and the corresponding projectile sizes) can be obtained using appropriate scaling (section \ref{sec:effects} below). It is important to point out that the post-impact focusing of the mass into two distinct locations, which has not been observed in previous simulations, is not due to the low velocity of the impacts, but rather the result of the specific target properties, namely their elongated shapes, the initial rotation and the relatively high porosity. All these properties are typical for real comets \citep{Lamy:2004lt}, but were not considered in previous impact studies. Moreover, the collision energies considered here are between the cratering and the catastrophic regime, an area that has not been well explored before. 

\subsection{Resulting final shapes}\label{sec:shapes}

A summary of our simulations, using various rotation rates, orientations as well as a range of impact points, is presented in Figs.  \ref{fig:shapes0.7} and  \ref{fig:shapes0.4}, which show the final shapes resulting from the collisions considered. We find that for the conditions investigated here, there is a reasonably large fraction of shapes that have 67P/C-G-like bi-lobe structures. While this is true for the two different targets (different axis ratio) considered here, bodies with two distinct lobes are more probable outcomes in collisions involving the more elongated target.

\subsection{Specific impact energies, heating and compaction} \label{sec:effects}
The specific impact energy is defined as
\begin{equation}\label{eq:qdef}
Q = 0.5 \mu_r V^2 / (M_t+M_p)
\end{equation}
where $\mu_r=M_p M_t /(M_t+M_p)$ is the reduced mass. We note that $\mu_r \simeq M_p$ for $M_p << M_t$, as it is the case here.
In Figure  \ref{fig:qcrit}, we compare the specific energies $Q_{sub}$ of  the sub-catastrophic impacts considered in this study to the specific impact energies for catastrophic collisions ($Q^*_D$) as well as for shape changing impacts on comet 67P/C-G  ($Q_{reshape}$) (see Paper I). Following the scaling law already applied for $Q^*_D$ and $Q_{reshape}$, we can write 
\begin{equation}\label{eq:qscaling}
Q_{sub} = a R^{3\mu}V^{2-3\mu}
\end{equation}
with $R$ = 2.52 km and $\mu$ = 0.42 (Paper I). The parameter $a$ is determined by the impact conditions used  for the two different targets (section \ref{sec:setup}). For the case with the  axis ratio of 0.4 with use the impact velocity of 250 m/s. Table \ref{table:scaling}  lists the values of parameter $a$ for  the various cases.

\begin{table}
\caption{Parameters (SI units) for the scaling law $Q_{crit} = a R^{3\mu}V^{2-3\mu}$, where $R$ is the target radius and $V$ the impact velocity. The scaling for $Q_{reshape}$ only holds for a fixed size, corresponding to comet 67P/C-G ($R$ = 1800 m).}            
\label{table:scaling}      
\centering                        
\begin{tabular}{c c c }        
\hline\hline              
Scaling & $\mu$ & $a$ \\    
\hline                        
   $Q^*_D$ & 0.42 & 4.00e-4 \\      
   $Q_{reshape}$ (average) & 0.42 & 2.50e-6 \\
   $Q_{sub}$ (0.4) & 0.42 & 1.66e-5 \\
   $Q_{sub}$ (0.7) & 0.42 & 4.90e-5 \\
   \hline                               
\end{tabular}
\end{table}

\subsection{Impact heating} \label{sec:heating}
\begin{figure}
\centering
\includegraphics[width=\hsize]{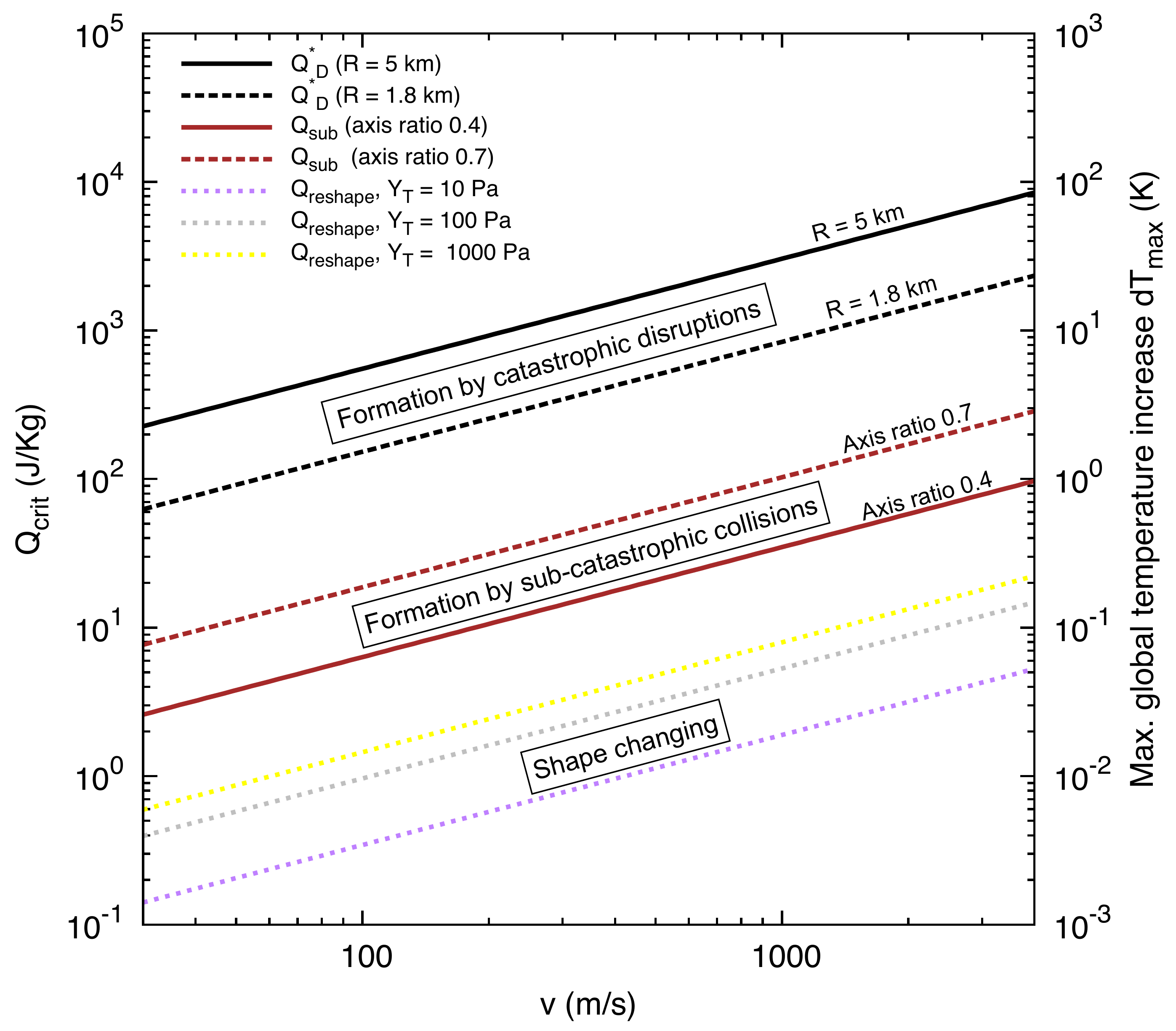}
\caption{Critical specific impact energies (adapted from Paper I): Shown are specific energies for catastrophic disruptions ($Q^*_D$), shape changing collisions ($Q_{reshape}$) and sub-catastrophic impacts ($Q_{sub}$). For the latter, a target radius $R$ = 2.52 km and impact velocities of $V$ = 300 m/s (axis ratio 0.7) and $V$ = 250 m/s (axis ratio 0.4) have been assumed.}
 \label{fig:qcrit}
 \end{figure}

Also shown in Figure \ref{fig:qcrit} is the maximal global temperature increase $dT$ resulting from the impacts. To estimate an upper limit for the global $dT$, we assumed that all kinetic impact energy is converted into internal energy: $dT=Q/c_p$ for which a constant heat capacity $c_p$ = 100 J/kg/K has been adopted (see Paper I).
This rough estimate already indicates that for impacts with energies comparable to $Q=Q_{sub}$, the maximal \emph{global} temperature increase must remain relatively small. The actual temperature distribution depends on how much of the available kinetic energy is converted into heating and what fraction of the heated material remains on the body. Moreover, due to the highly dissipative characteristics of porous material, the compressed and therefore heated region remains very localised to the vicinity of the impact and therefore such an impact affects little the bulk content of volatile elements and the bulk porosity. 

\begin{figure}
\centering
\includegraphics[width=\hsize]{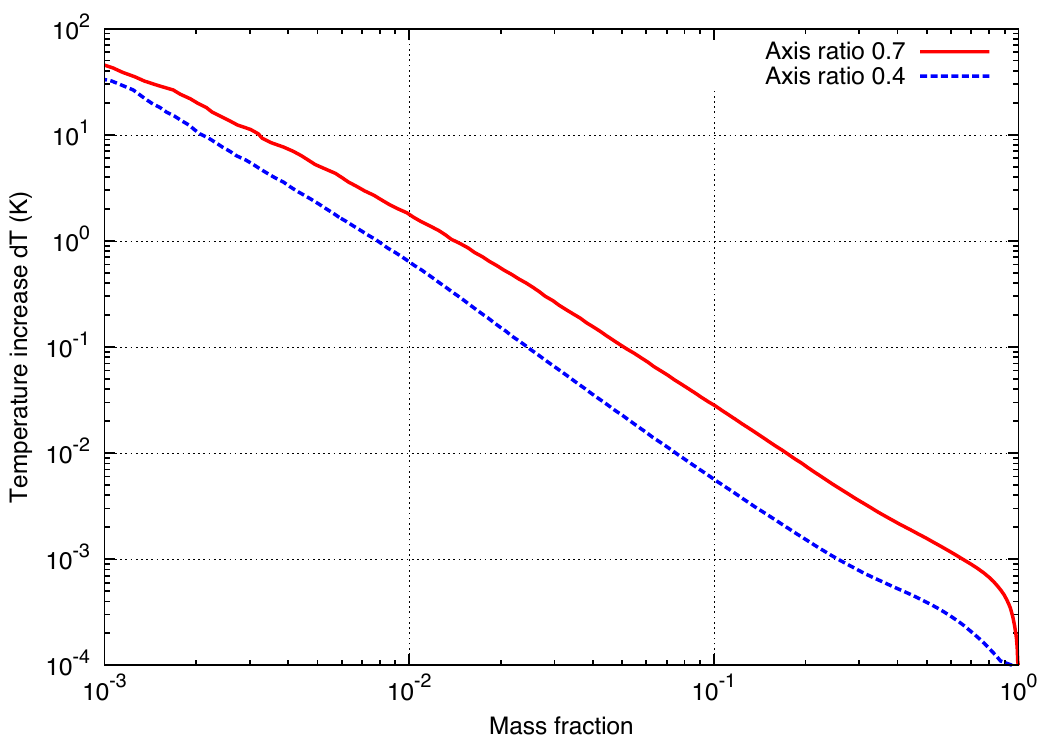}
\caption{Fraction of material in the final body that experienced a temperature increase larger than a certain value $dT$. The curves correspond to the cases shown in Figure   \ref{fig:snapshots0.7} (axis ratio 0.7) and Figure \ref{fig:snapshots0.4} (axis ratio 0.4), respectively.}
 \label{fig:u_frac}
 \end{figure}

We demonstrate this by plotting in Figure  \ref{fig:u_frac} the fraction of material that actually experienced a temperature increase larger than a certain $dT$ for the two cases presented in Figures \ref{fig:snapshots0.7} and \ref{fig:snapshots0.4}. As it can be seen, only $\lesssim$ 1\% of the mass in the final bodies experienced a temperature increase larger than a few K. According to the scaling laws (Figure \ref{fig:qcrit}) higher impact velocities would require slightly higher specific impact energies in order to result in a similar final bi-lobed configuration. For km/s impact velocities, we would therefore expect the curves in figure  \ref{fig:u_frac} to be slightly shifted towards higher $dT$. However, as discussed above, the heating would remain localised and on a global scale, $dT$ remains limited to relatively small values even at high impact speeds. We note that the bulk of the $dT$ values found here are generally much smaller than the sublimation temperatures of the observed super-volatiles, which are are typically $T_s >$ 20 K \citep{Yamamoto:1985,Meech:2004ms}.

\subsection{Porosity evolution} \label{sec:porosity}
\begin{figure}
\centering
\includegraphics[width=\hsize]{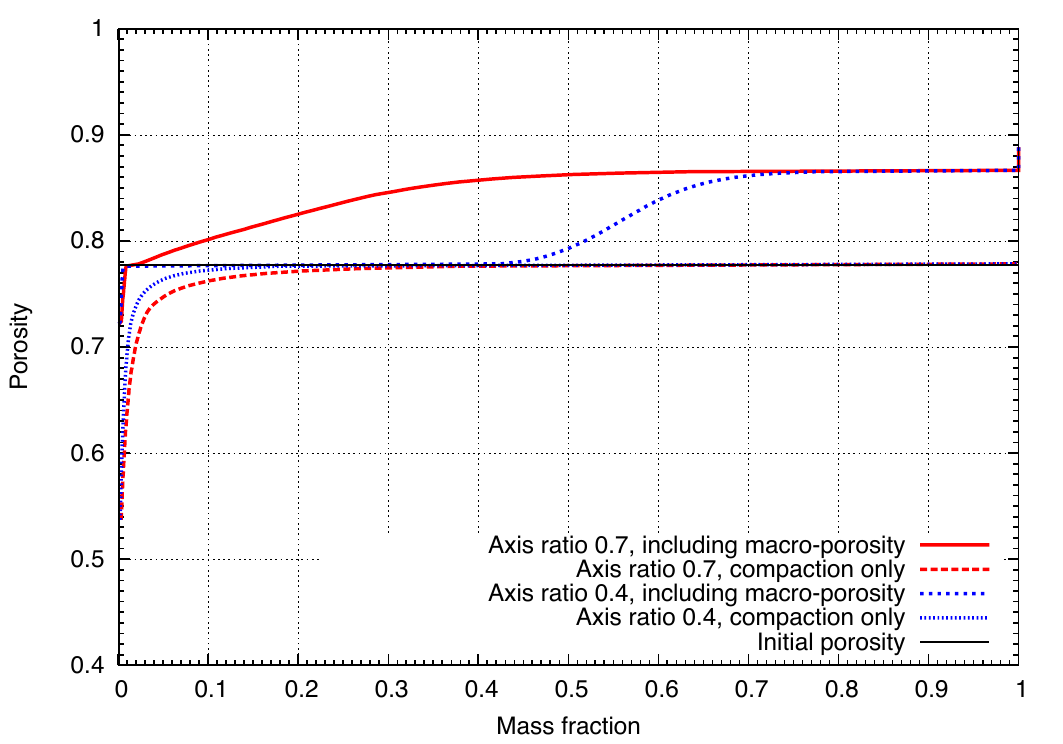}
\caption{Cumulative distribution of the porosity in the final body. The curves correspond to the cases shown in Figure   \ref{fig:snapshots0.7} (axis ratio 0.7) and Figure \ref{fig:snapshots0.4} (axis ratio 0.4), respectively. The porosity calculation takes into account compaction as well as the increase of macorporosity.  For comparison, the porosity distributions resulting from compaction only are shown as well. The final average porosity (compaction plus addition of macroporosity by reaccumulation) is 85.1\% (axis ratio 0.7) and 82.3\% (axis ratio 0.4), while the initial porosity was 77.8\%.}
 \label{fig:dist_frac}
 \end{figure}

Using the same procedure as in Paper I, we  compute the cumulative distribution of the porosity in the final body, which takes into account compaction as well as the addition of macroporosity by reaccumulation of ejected material (Figure  \ref{fig:dist_frac}).

Only a small fraction of the target mass ($\lesssim$ 5\%) significantly compacted (the porosity is reduced by $\gtrsim$ 10\%). On the other hand, a considerable amount of material experiences ejection and subsequent reaccumulation, thereby the introduction of macroporosity.  Consequently, the final average porosities are slightly higher than the initial porosity (Figure  \ref{fig:dist_frac}).

\begin{figure}
\centering
\includegraphics[width=\hsize]{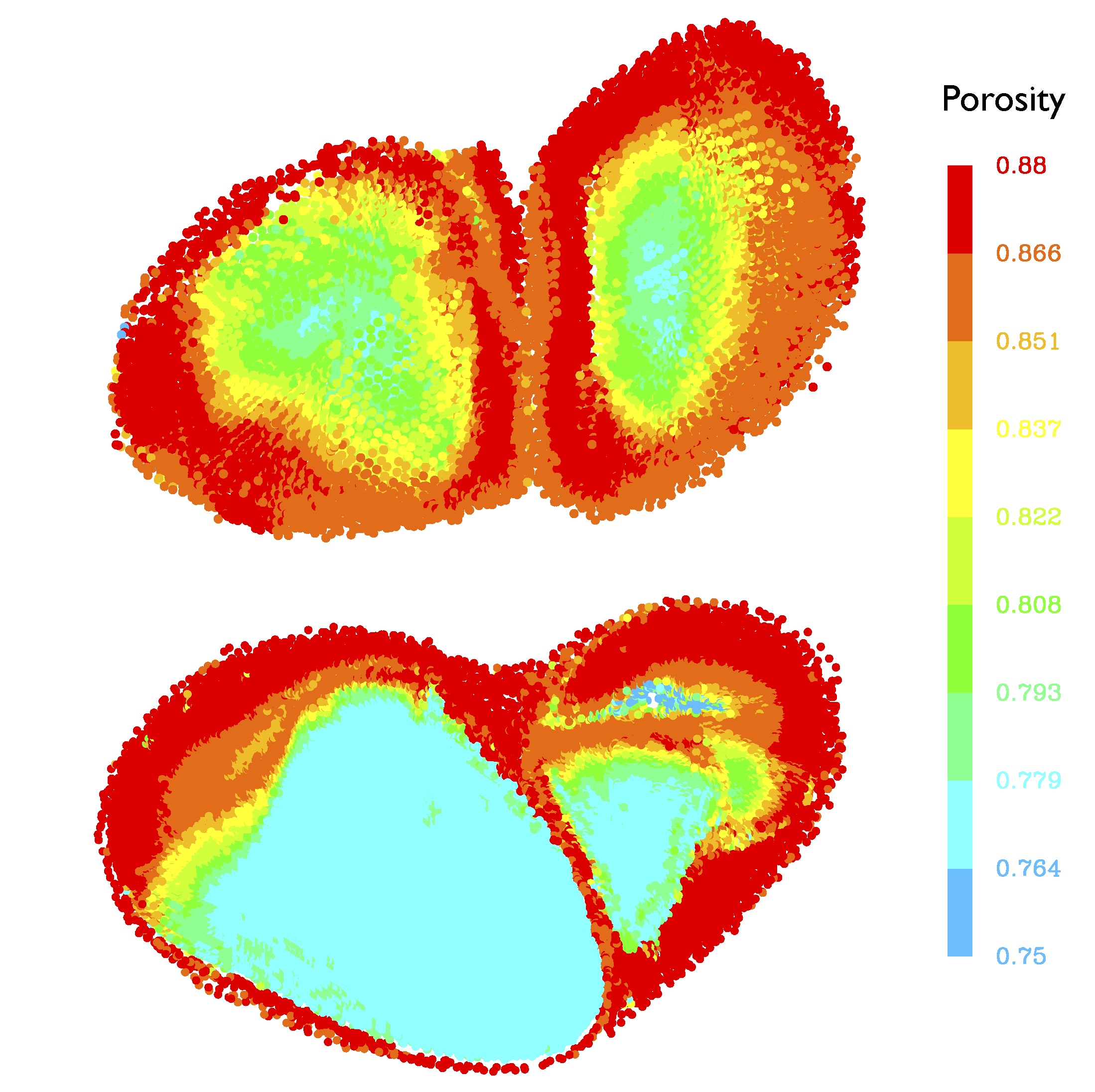}
\caption{Cross-sections showing the distribution of the porosity in the final body. The plots correspond to the cases shown in Figure   \ref{fig:snapshots0.7} (top) and Figure \ref{fig:snapshots0.4} (bottom).}
 \label{fig:dist_frac_cross}
 \end{figure}

Figure \ref{fig:dist_frac_cross} displays cross-sections (in the x-y-plane) of the two bodies resulting from the impact simulations, showing the distribution of the final porosity. The porosity variations are generally fairly small, consistent with the observation that the interior of the nucleus is homogeneous on a global scale \citep{Paetzold:2016pa}.  On a large scale, porosity slightly decreases with depth.  On smaller scales, layers of varying porosities are observed, suggesting that stratification such as observed on 67P/C-G \citep{Massironi:2015ma}, may be produced by the reaccumulation of material on each single lobe (see also Figures \ref{fig:snapshots0.7} and \ref{fig:snapshots0.4}). 
In some locations, porosity decreases with increasing depths, as suggested by CONSERT measurements of the first $\sim$ 100 m of the subsurface \citep{Ciarletti:2015cl}. However, we stress that the spatial resolution of our simulations (of the order of $\sim$ 100 m) does not allow to directly compare our results with these measurements. It also prevents us from making predictions regarding the size distribution of reaccumulated boulders.

\section{Survival probabilities}\label{sec:probability}

\subsection{Number of collisions}
If the currently observed 67P/C-G structure did indeed form as a result of a collision event as shown above, for it to be observed today implies that it has not suffered any shape-changing collisions since the time of its formation. 

Assuming that the formation of a bi-lobed structure of the size of 67P/C-G requires at least a specific impact energy of  $Q_{form}$, we can estimate the average number of subsequent shape-changing collisions which have the minimum energy $Q_{reshape}$ (as determined in Paper I), as well as the probability to avoid all of these collisions. 
To allow for this, we calculate from Equation \ref{eq:qdef} the minimal projectile radius delivering a specific impact energy $Q \ge Q_{min}$ as 
\begin{equation}
R_{min} = (2 Q_{min} V)^{1/3} R_t / V 
\end{equation}
where we assume that target and impactor have the same density and  that $M_p << M_t$ (for impact velocities of a few hundred m/s, this is true even for $Q_{min}=Q^*_D$). The number of impacts on a target of size $R_t$ by projectiles within in the size range $R_{min} \leq R_t \leq R_{max}$ during a time interval $\delta t$, can be written as \citep[e.g. ][]{Morbidelli:2015vm}:
\begin{equation}\label{eq:ncoll}
N = P_i \delta t \int_{R_{min}}^{R_{max}} \pi (R_t+R_p )^2 N_p(R_p )dR_p 
\end{equation}
where $P_i$ is the average intrinsic collision probability. Following \citet{Morbidelli:2015vm} we use $R_{max}$ = 50 km. $N_p(R_p)$ represents the number of projectiles with a radius between $R_p$ and $R_p + dR_p$. This distribution is not known precisely but in agreement with other studies of small body size distribution, we assume a differential size distribution of bodies given by $dN/dr \sim r^q$. For the exponent $q$ we use the same values  as in calculations performed in Paper I: $q=-2.5, -3.0$ and $-3.5$.

Equation \ref{eq:ncoll} can be used to compute the number of collisions $N(Q_{min}=Q_{form})$ within the time interval $\delta t$ having a minimal specific impact energy $Q_{form}$ allowing the formation of 67P/C-G-like structure. In addition to these collisions, during the same time interval a number of reshaping collisions $N(Q_{min}=Q_{reshape})$ will take place. Given that $Q_{form}$ is larger than $Q_{reshape}$, $N(Q_{min}=Q_{reshape})$ will be larger than $N(Q_{min}=Q_{form})$. In other words, for one formation event ($N(Q_{min}=Q_{form})=1$) there will be on average $N_{rs,norm}$ reshaping collisions. This can be expressed mathematically as: 

\begin{equation}
N_{rs,norm}=\frac{N(Q_{min}=Q_{reshape})-N(Q_{min}=Q_{form})}{N(Q_{min}=Q_{form})},
\end{equation}
or rewritten:
\begin{equation}
N_{rs,norm}=\frac{N(Q_{min}=Q_{reshape})}{N(Q_{min}=Q_{form})}-1
\end{equation}
We compute $N_{rs,norm}$ for the scenario of the formation of the 67P/C-G shape as a result of either a catastrophic break-up event (where we assume that a specific energy of at least $Q_{form}=Q^*_D$ is required) or a sub-catastrophic formation event as illustrated above (where $Q_{form}=Q_{sub}$). The numbers obtained are shown in Table  \ref{table:ncoll}  for two different projectile velocities $V$ = 500 m/s and $V$ = 2 km/s, which are representative for the relative velocities of small bodies within in the planetesimal disc before and after dispersal, respectively  \citep{Morbidelli:2015vm}. 

\begin{table}
\caption{Average number of shape-changing collisions $N_{rs,norm}$ for one formation event [$N(Q_{min}=Q_{form})$ = 1]. $N_{rs,norm}$ is computed for the catastrophic break-up with $Q_{form}=Q^*_D$ (with $R$ = 3 km) and sub-catastrophic formation with $Q_{form} = Q_{sub}$ and two axis ratios (0.4 and 0.7, respectively).}      
\label{table:ncoll}
\centering                        
\begin{tabular}{c l c c c}        
\hline\hline              
 $V$  & Formation type &$q$ = -2.5 & $q$ = -3 & $q$ = -3.5 \\
   \hline 
0.5 &Catastrophic&4.60&23.1&90.9 \\
km/s &Sub-catastrophic (0.7)&2.09&4.93&9.2 \\
&Sub-catastrophic (0.4)&1.12&2.18&3.46 \\
  \hline 
2.0 &Catastrophic&6.58&30.1&112\\
km/s &Sub-catastrophic (0.7)&2.57&5.41&9.9\\
&Sub-catastrophic (0.4)&1.34&2.30&3.61\\
   \hline                               
\end{tabular}
\end{table}

\subsection{Probabilities to avoid sub-sequent collisions}
To compute the survival probability of a 67P/C-G-like structure formed by a collision event of a certain type, we can use the number of reshaping collisions per unit number of formation collisions $N_{rs,norm}$ derived in the previous section. In the following, we estimate this survival probability $P_{survival}$ for the different formation scenarios by assuming that a 67P/C-G structure was formed as a result of the last possible collision. This means the last collision involving the required specific energy (e.g., $Q_{form}=Q^*_D$ or $Q_{form}=Q_{sub}$). We then take into account that on average, only a fraction $f$ of the shape changing collisions $N_{rs,norm}$ take place \emph{after} the structure formed:

\begin{equation}
n_f = f N_{rs,norm}
\end{equation}
We further assume that these $n_f$ events follow a Poisson distribution. In this case, the probability that all subsequent shape changing collisions are avoided is given by
\begin{equation}
P(f)=e^{-n_f}=e^{-f N_{rs,norm}}
\end{equation}
To obtain the total survival probability $P_{survival}$ we integrate over all possible orders in which the collisions may take place
\begin{equation}
P_{survival}=\int_0^1 \frac{P(f)}{N_{rs,norm}+1} df 
\end{equation}
where we use the weight factor $1/(N_{rs,norm}+1)$. By construction, $P_{survival}=1$ for $N_{rs,norm}=0$.

\begin{table}
\caption{Survival probabilities in the different scenarios. Computed is the bi-lobe shape survival probability, $P_{survival}$, against any subsequent shape changing collision. These survival probabilities are computed for three different scenarios a) a catastrophic disruption of a target with $R_t$ = 3 km, b) a sub-catastrophic impact (for two target axis ratios 0.7 and 0.4), and c) a primordial formation. See text for details. For the catastrophic disruption,  the probability computation in the second row takes into account the reduced abundance of $R_t$ = 3 km bodies with respect to 67P/C-G sized bodies with $R_t$ $\sim$ 1.8 km (a factor of $c_n$ = 0.29 for $q$ = -2.5, $c_n$ = 0.22 for $q$ = -3 and $c_n$ = 0.17 for $q$ = -3.5, respectively).}      
\label{table:psurvival}
\centering                        
\begin{tabular}{l c c c}        
\hline\hline              
Type &$q$ = -2.5 & $q$ = -3 & $q$ = -3.5 \\
\hline
 Catastrophic (3 km) &2.00E-02&1.07E-03&7.85E-05\\
Including $c_n$ &1.52E-03&5.03E-05&2.25E-06\\
\hdashline 
\hdashline
 Sub-catastrophic (0.7) &1.01E-01&2.87E-02&9.30E-03 \\
Sub-catastrophic (0.4)& 2.35E-01&1.18E-01&5.85E-02 \\
\hdashline 
\hdashline 
Primordial formation:  & & & \\
$T_{disc}$= 400 Myrs & < 1e-28 & < 1e-90 & < 1e-228\\
$T_{disc}$= 0 Myrs & 4.69E-02&1.42E-08&2.42E-49\\
\hline                               
\end{tabular}
\end{table}

\begin{figure*}
\centering
\includegraphics[width=\hsize]{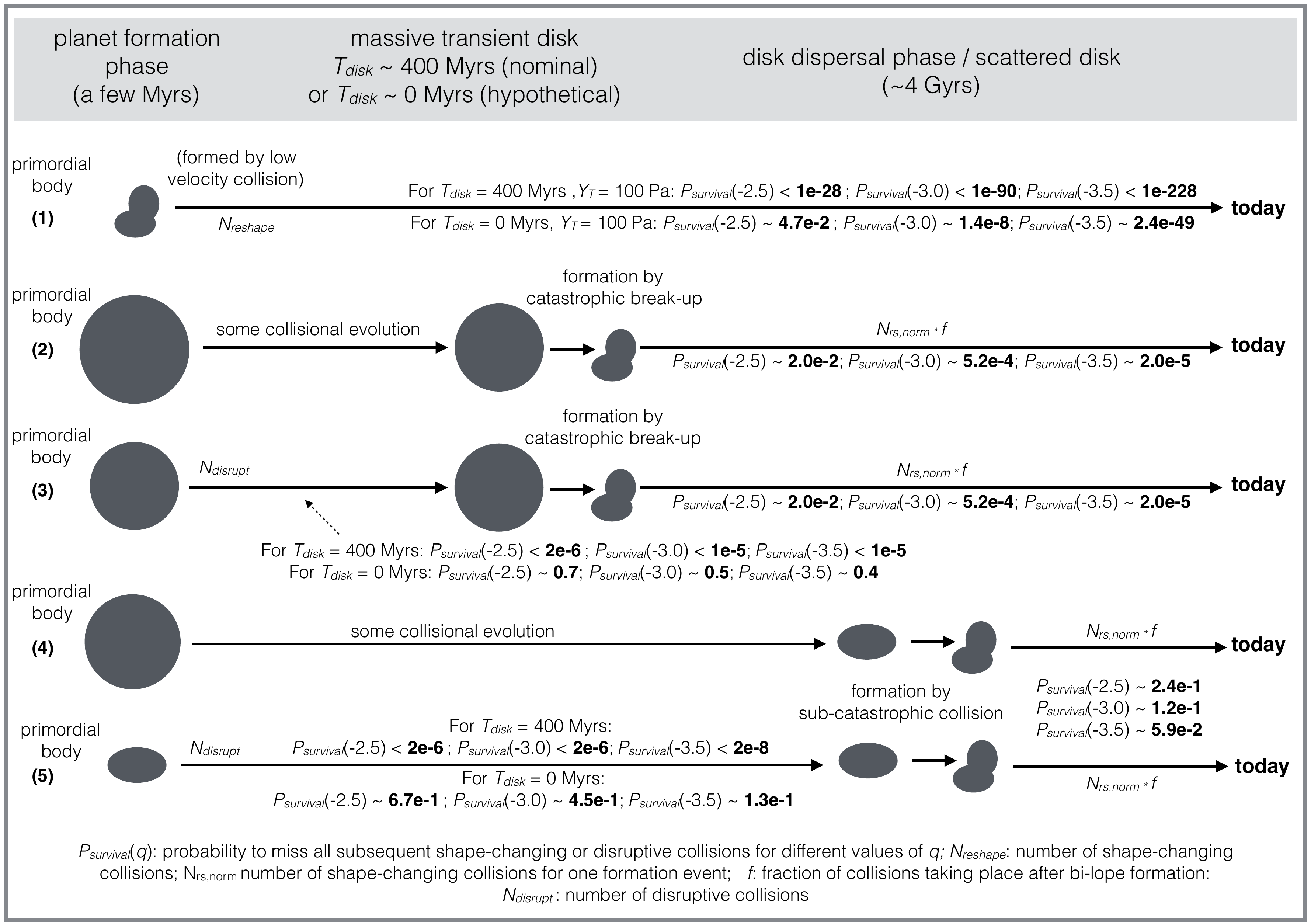}
\caption{Scenarios for the formation of a 67P/C-G-like bi-lobe structure. For each scenario, the probability to avoid all subsequent shape-changing collisions is shown. For the cases of a late formation by a collision, we also indicate the probability that the parent body did not experience any prior catastrophic disruption.}
 \label{fig:scenarios}
 \end{figure*}

The results of these calculations\footnote{Note that the calculations presented here do not take into account the probability of the formation of a bi-lobe shape in a specific impact event with a given energy. As discussed in section \ref{sec:shapes}, this probability is reasonably high, but is not possible to quantify.} are given in Table \ref{table:psurvival}. For comparison, we also show in Table  \ref{table:psurvival} the survival probabilities in the case of a primordial formation of a 67P/C-G-like bi-lobed structure (Paper I). For the standard scenario with a giant planet instability taking place at $\sim$ 400 Myrs, we adapt the number of disruptive collisions computed in \citet{Morbidelli:2015vm} to the number of shape-changing collisions using Equation  \ref{eq:ncoll}  and $Q_{reshape}$ (for a tensile strength of 100 Pa; Paper I). The survival probability then corresponds to the probability of missing all shape-changing collisions. For the scenario with no transient disc, the survival probabilities are computed in Paper I. 

The results vary significantly between the different scenarios, and also strongly depend upon the size distribution of available projectiles (characterised by $q$). Preliminary results from the New Horizons mission to Pluto and Charon based on the crater size frequency distribution suggest that $q\sim-3.3$ \citep{Singer:2015ss}. We note that based on the most recent results it has been suggested that there may be a deficit of small objects \citep{Singer:2016sm}; see discussion in Paper I, section 4. For size distributions characterised by $q \ge$ 3, we find that  $P_{survival}$ is of the order of 1-10 \% in the framework of a formation following a sub-catastrophic collision while $P_{survival} <$ 0.1 \% in case of the formation following directly a catastrophic disruption. In the case of a primordial formation, the survival probability $P_{survival}<10^{-8}$, even in the conservative assumption that there was no initial massive transient planetesimal disc. We note that alternative models of the dynamical evolution \citep{Davidsson:2016ds} (see discussion in Paper I) predict a much smaller collisional evolution and prefer shallower slopes of the size distribution, which would generally increase the survival probabilities computed here.
An overview of the various scenarios considered in our study is presented in Figure \ref{fig:scenarios}. 

Note that so far we have not taken into account the possibility of a final shape evolution via a fission-merging cycle \citep{Scheeres:2016sh}. In this scenario, the bi-lobe forming  collisions presented here would not form 67P/C-G directly, but rather a two-component structure with the right size ratio. This two-component body would then later evolve into the final shape via (one or several) fission-merging cycle, once the comet enters the inner solar system. In this scenario, the survival probabilities might be somewhat larger than those computed above since general two-component structures require a slightly higher impact energy to be 'destroyed' (Paper I).

It is likely that the parent body from which the 67P/C-G-like structure formed is not primordial itself (as indicated by the survival probabilities in Figure  \ref{fig:scenarios})  and has already experienced some collisional evolution or maybe a result of a collisional cascade.

\section{Discussion and Conclusions}\label{conclusions}

The analysis of the survival of the global structure of comet 67P/C-G shape (Paper I) strongly suggests that such a shape cannot be primordial. It must have formed as a result of a collision at a subsequent time (most probably within the last Gyr). At this time, the relative velocities between cometary-sized bodies are such ($V>>V_{esc}$) that the formation mechanism invoked previously, namely the collisional mergers at low velocity ($V \approx V_{esc}$) of similar-sized bodies \citep{Jutzi:2015ja}, cannot work directly anymore. 

In this paper, we present an alternative scenario. We investigate the final shapes resulting from a new type of low-energy, sub-catastrophic impacts on elongated, rotating bodies, using a 3D SPH shock physics code. Our modelling results suggest that such collisions result in "splitting" events which frequently lead to formation of  bi-lobe 67P/C-G-like shapes. This mechanism might not only explain the bi-lobe shape of some cometary nuclei but could potentially also provide an explanation for structures observed in the asteroid population, as for instance the particular forms of the asteroids 25143 Itokawa or 4179 Toutatis.

According to our model, comets are not primordial in the sense that their shape and structure formed during the initial stages of the formation of the Solar System. Rather, the final structure is the result of the last major shape-forming impact. The sub-catastrophic collisions investigated here provide a possibility of bi-lobe formation with small impact energies. Such small-scale impacts are much more frequent than catastrophic disruptions and the probability for such a shape-forming event to occur without a subsequent shape-destroying event occurring until today is estimated to be reasonably high.
We note that the two-component structure resulting from the type of collisions investigated here might further evolve by fission-merging cycle once the comet enters the inner Solar system, as suggested recently \citep{Scheeres:2016sh}.

 Although the individual collisions considered in this work can alter the global shape, their respective energy is small enough not to lead individually to any substantial global scale heating or compaction. In this sense, our formation model is consistent with the observed "pristinity" of 67P/C-G \citep[e.g. ][]{Rubin:2015ra,Bieler:2015ba}. However, it is likely that the parent body from which the 67P/C-G-like structure ultimately formed must have undergone significant collisional evolution (Figure  \ref{fig:scenarios}; see also Paper I), or is itself the result of a collisional disruption of a larger parent body. Several paths through the collisional cascade being able to lead to the similar-sized bodies, the cumulative effects of impact heating and compaction experienced during the 4.6 Gyrs of evolution by the material components that eventually form the comets observed today are difficult to establish with certainty. This formation degeneracy implies that it is not possible to reconstruct uniquely the detailed collision history nor the number and sizes of the parent bodies. Nevertheless, an upper limit for the size of parent bodies is given by the fact that larger bodies are subjected to internal heating by short-lived radionuclides \citep[e.g.][]{Prialnik:2008ps} that will significantly alter the pristine nature of the material and therefore be incompatible with observations of porosity and content in highly volatile elements in comets. Interestingly, this upper limit of the size of the parent body coupled with the requirement that the cumulative effects of impacts in terms of compaction and volatile losses can only be very moderate provide strong constraints for the duration and intensity of the collisional bombardment. 
 
Whether these constraints are compatible with a scenario of a massive planetesimal disc phase existing for 450 Myr, as proposed by the Nice model \citep{Tsiganis:2005tg,Gomes:2005gl,Morbidelli:2012ms}, remains to be analysed in detail. We note that \citet[][]{Davidsson:2016ds} suggest that the number of objects in the disc was much smaller, leading to less collisions (however, this model has other issues; see discussion in Paper I). On the other hand, our analysis of heating and porosity evolution in the impacts considered here as well the regimes investigated in Paper I (shape-changing impacts as well as  catastrophic disruptions) indicates that collisionally processed objects may still look "primitive". It is found that such bodies can still have a high porosity, and could have retained their volatiles, since these collisions generally do not lead to large-scale heating of the material bound in the largest remnant. A more detailed study of the outcome of large-scale catastrophic disruptions is currently in progress (Schwartz et al., in prep., 2016). 

In any case, given our current understanding of the dynamics of the small bodies in the outer solar system, it is unlikely that the currently observed shape of comet 67P/C-G is primordial (even in the hypothetical scenario in which no initial massive planetesimal disc was existing; Paper I).  According to the calculations presented here, it may have formed as a result of the last major shape-forming impact. Nevertheless, should future investigations show that the collisional cascade does not preserve the pristine nature of cometary material, we would be facing the conclusion that the current knowledge of the dynamics of small bodies in the outer regions of the solar system is seriously flawed. In this sense, comets provide invaluable tools to probe the origin and evolution of our solar system.

\begin{acknowledgements}
     M.J. and W.B. acknowledge support from the Swiss NCCR PlanetS. We thank the referees B. Davidsson and N. Movshovitz for their thorough review which helped to improve the paper substantially.

 \end{acknowledgements}

%
\bibliographystyle{aa} 
\bibliography{../../paper_disruption/bibdata} 

\begin{thebibliography}{35}
\expandafter\ifx\csname natexlab\endcsname\relax\def\natexlab#1{#1}\fi

\bibitem[{Benz \& Asphaug(1995)}]{Benz:1995hx}
Benz, W. \& Asphaug, E. 1995, Computer Physics Communications, 87, 253

\bibitem[{{Bieler} {et~al.}(2015){Bieler}, {Altwegg}, {Balsiger},
  {et~al.}}]{Bieler:2015ba}
{Bieler}, A., {Altwegg}, K., {Balsiger}, H., {et~al.} 2015, Nature, 526, 678

\bibitem[{{Ciarletti} {et~al.}(2015){Ciarletti}, {Levasseur-Regourd}, {Lasue},
  {Statz}, {Plettemeier}, {H{\'e}rique}, {Rogez}, \&
  {Kofman}}]{Ciarletti:2015cl}
{Ciarletti}, V., {Levasseur-Regourd}, A.~C., {Lasue}, J., {et~al.} 2015, \aap,
  583, A40

\bibitem[{{Davidsson} {et~al.}(2016){Davidsson}, {Sierks}, {G{\"u}ttler},
  {Marzari}, {Pajola}, {Rickman}, {A'Hearn}, {Auger}, {El-Maarry}, {Fornasier},
  {Guti{\'e}rrez}, {Keller}, {Massironi}, {Snodgrass}, {Vincent}, {Barbieri},
  {Lamy}, {Rodrigo}, {Koschny}, {Barucci}, {Bertaux}, {Bertini}, {Cremonese},
  {Da Deppo}, {Debei}, {De Cecco}, {Feller}, {Fulle}, {Groussin}, {Hviid},
  {H{\"o}fner}, {Ip}, {Jorda}, {Knollenberg}, {Kovacs}, {Kramm}, {K{\"u}hrt},
  {K{\"u}ppers}, {La Forgia}, {Lara}, {Lazzarin}, {Lopez Moreno},
  {Moissl-Fraund}, {Mottola}, {Naletto}, {Oklay}, {Thomas}, \&
  {Tubiana}}]{Davidsson:2016ds}
{Davidsson}, B.~J.~R., {Sierks}, H., {G{\"u}ttler}, C., {et~al.} 2016, \aap,
  592, A63

\bibitem[{{Gomes} {et~al.}(2005){Gomes}, {Levison}, {Tsiganis}, \&
  {Morbidelli}}]{Gomes:2005gl}
{Gomes}, R., {Levison}, H.~F., {Tsiganis}, K., \& {Morbidelli}, A. 2005, \nat,
  435, 466

\bibitem[{{H{\"a}ssig} {et~al.}(2015){H{\"a}ssig}, {Altwegg}, {Balsiger}, \&
  {etal.}}]{Haessig:2015he}
{H{\"a}ssig}, M., {Altwegg}, K., {Balsiger}, H., \& {etal.} 2015, Science, 347

\bibitem[{Jutzi(2015)}]{Jutzi:2015gb}
Jutzi, M. 2015, Planetary and Space Science, 107, 3

\bibitem[{Jutzi \& Asphaug(2015)}]{Jutzi:2015ja}
Jutzi, M. \& Asphaug, E. 2015, Science, 348, 1

\bibitem[{Jutzi {et~al.}(2008)Jutzi, Benz, \& Michel}]{Jutzi:2008kp}
Jutzi, M., Benz, W., \& Michel, P. 2008, Icarus, 198, 242

\bibitem[{Jutzi {et~al.}(2015)Jutzi, Holsapple, W{\"u}nnemann, \&
  Michel}]{Jutzi:2015ux}
Jutzi, M., Holsapple, K.~A., W{\"u}nnemann, K., \& Michel, P. 2015, Asteroids
  IV, 1

\bibitem[{{Lamy} {et~al.}(2004){Lamy}, {Toth}, {Fernandez}, \&
  {Weaver}}]{Lamy:2004lt}
{Lamy}, P.~L., {Toth}, I., {Fernandez}, Y.~R., \& {Weaver}, H.~A. 2004, {The
  sizes, shapes, albedos, and colors of cometary nuclei}, ed. M.~C. {Festou},
  H.~U. {Keller}, \& H.~A. {Weaver}, 223--264

\bibitem[{{Le Roy} {et~al.}(2015){Le Roy}, {Altwegg}, {Balsiger}, {Berthelier},
  {Bieler}, {Briois}, {Calmonte}, {Combi}, {De Keyser}, {Dhooghe}, {Fiethe},
  {Fuselier}, {Gasc}, {Gombosi}, {H{\"a}ssig}, {J{\"a}ckel}, {Rubin}, \&
  {Tzou}}]{LeRoy:2015ra}
{Le Roy}, L., {Altwegg}, K., {Balsiger}, H., {et~al.} 2015, \aap, 583, A1

\bibitem[{{Marchi} {et~al.}(2015){Marchi}, {Rickman}, {Massironi}, {Marzari},
  {El-Maari}, {Besse}, {Thomas}, {Barbieri}, {Barucci}, {Fornasier},
  {Giacomini}, {Keller}, {Kuehrt}, {Lamy}, {Lazzarin}, {Mottola}, {Naletto},
  {Pajola}, \& {Sierks}}]{Marchi:2015mr}
{Marchi}, S., {Rickman}, H., {Massironi}, M., {et~al.} 2015, in Lunar and
  Planetary Science Conference, Vol.~46, Lunar and Planetary Science
  Conference, 1532

\bibitem[{{Massironi} {et~al.}(2015){Massironi}, {Simioni}, {Marzari},
  {Cremonese}, {Giacomini}, {Pajola}, {Jorda}, {Naletto}, {Lowry}, {El-Maarry},
  {Preusker}, {Scholten}, {Sierks}, {Barbieri}, {Lamy}, {Rodrigo}, {Koschny},
  {Rickman}, {Keller}, {A'Hearn}, {Agarwal}, {Auger}, {Barucci}, {Bertaux},
  {Bertini}, {Besse}, {Bodewits}, {Capanna}, {da Deppo}, {Davidsson}, {Debei},
  {de Cecco}, {Ferri}, {Fornasier}, {Fulle}, {Gaskell}, {Groussin},
  {Guti{\'e}rrez}, {G{\"u}ttler}, {Hviid}, {Ip}, {Knollenberg}, {Kovacs},
  {Kramm}, {K{\"u}hrt}, {K{\"u}ppers}, {La Forgia}, {Lara}, {Lazzarin}, {Lin},
  {Lopez Moreno}, {Magrin}, {Michalik}, {Mottola}, {Oklay}, {Pommerol},
  {Thomas}, {Tubiana}, \& {Vincent}}]{Massironi:2015ma}
{Massironi}, M., {Simioni}, E., {Marzari}, F., {et~al.} 2015, \nat, 526, 402

\bibitem[{{Meech} \& {Svoren}(2004)}]{Meech:2004ms}
{Meech}, K.~J. \& {Svoren}, J. 2004, {Using cometary activity to trace the
  physical and chemical evolution of cometary nuclei}, ed. G.~W. {Kronk},
  317--335

\bibitem[{{Melosh}(1989)}]{Melosh:1989}
{Melosh}, H.~J. 1989, {Impact cratering: A geologic process}

\bibitem[{{Michel} {et~al.}(2003){Michel}, {Benz}, \&
  {Richardson}}]{Michel:2003mb}
{Michel}, P., {Benz}, W., \& {Richardson}, D.~C. 2003, \nat, 421, 608

\bibitem[{{Michel} {et~al.}(2001){Michel}, {Benz}, {Tanga}, \&
  {Richardson}}]{Michel:2001mb}
{Michel}, P., {Benz}, W., {Tanga}, P., \& {Richardson}, D.~C. 2001, Science,
  294, 1696

\bibitem[{{Michel} \& {Richardson}(2013)}]{Michel:2013mr}
{Michel}, P. \& {Richardson}, D.~C. 2013, \aap, 554, L1

\bibitem[{{Morbidelli} {et~al.}(2012){Morbidelli}, {Marchi}, {Bottke}, \&
  {Kring}}]{Morbidelli:2012ms}
{Morbidelli}, A., {Marchi}, S., {Bottke}, W.~F., \& {Kring}, D.~A. 2012, Earth
  and Planetary Science Letters, 355, 144

\bibitem[{{Morbidelli} \& {Rickman}(2015)}]{Morbidelli:2015vm}
{Morbidelli}, A. \& {Rickman}, H. 2015, \aap, 583, A43

\bibitem[{Mumma {et~al.}(1993)Mumma, Weissman, \& Stern}]{Mumma:1993mw}
Mumma, M.~J., Weissman, P.~R., \& Stern, S.~A. 1993, In: Protostars and planets
  III (A93-42937 17-90), 1177

\bibitem[{Nyffeler(2004)}]{Nyffeler:2004tz}
Nyffeler, B. 2004, PhD thesis, University of Bern

\bibitem[{{P{\"a}tzold} {et~al.}(2016){P{\"a}tzold}, {Andert}, {Hahn}, {Asmar},
  {Barriot}, {Bird}, {H{\"a}usler}, {Peter}, {Tellmann}, {Gr{\"u}n},
  {Weissman}, {Sierks}, {Jorda}, {Gaskell}, {Preusker}, \&
  {Scholten}}]{Paetzold:2016pa}
{P{\"a}tzold}, M., {Andert}, T., {Hahn}, M., {et~al.} 2016, \nat, 530, 63

\bibitem[{{Prialnik} {et~al.}(2008){Prialnik}, {Sarid}, {Rosenberg}, \&
  {Merk}}]{Prialnik:2008ps}
{Prialnik}, D., {Sarid}, G., {Rosenberg}, E.~D., \& {Merk}, R. 2008, \ssr, 138,
  147

\bibitem[{{Rickman} {et~al.}(2015){Rickman}, {Marchi}, {A'Hearn}, {Barbieri},
  {El-Maarry}, {G{\"u}ttler}, {Ip}, {Keller}, {Lamy}, {Marzari}, {Massironi},
  {Naletto}, {Pajola}, {Sierks}, {Koschny}, {Rodrigo}, {Barucci}, {Bertaux},
  {Bertini}, {Cremonese}, {Da Deppo}, {Debei}, {De Cecco}, {Fornasier},
  {Fulle}, {Groussin}, {Guti{\'e}rrez}, {Hviid}, {Jorda}, {Knollenberg},
  {Kramm}, {K{\"u}hrt}, {K{\"u}ppers}, {Lara}, {Lazzarin}, {Lopez Moreno},
  {Michalik}, {Sabau}, {Thomas}, {Vincent}, \& {Wenzel}}]{Rickman:2015wu}
{Rickman}, H., {Marchi}, S., {A'Hearn}, M.~F., {et~al.} 2015, \aap, 583, A44

\bibitem[{{Rubin} {et~al.}(2015){Rubin}, {Altwegg}, {Balsiger}, {Bar-Nun},
  {Berthelier}, {Bieler}, {Bochsler}, {Briois}, {Calmonte}, {Combi}, {De
  Keyser}, {Dhooghe}, {Eberhardt}, {Fiethe}, {Fuselier}, {Gasc}, {Gombosi},
  {Hansen}, {H{\"a}ssig}, {J{\"a}ckel}, {Kopp}, {Korth}, {Le Roy}, {Mall},
  {Marty}, {Mousis}, {Owen}, {R{\`e}me}, {S{\'e}mon}, {Tzou}, {Waite}, \&
  {Wurz}}]{Rubin:2015ra}
{Rubin}, M., {Altwegg}, K., {Balsiger}, H., {et~al.} 2015, Science, 348, 232

\bibitem[{{Scheeres} {et~al.}(2016){Scheeres}, {Hirabayashi}, {Chesley},
  {Marchi}, {McMahon}, {Steckloff}, {Mottola}, {Naidu}, \&
  {Bowling}}]{Scheeres:2016sh}
{Scheeres}, D.~J., {Hirabayashi}, T., {Chesley}, S., {et~al.} 2016, in Lunar
  and Planetary Science Conference, Vol.~47, Lunar and Planetary Science
  Conference, 1615

\bibitem[{{Sierks} {et~al.}(2015){Sierks}, {Barbieri}, {Lamy}, {Rodrigo},
  {Koschny}, {Rickman}, {Keller}, {Agarwal}, {A'Hearn}, {Angrilli}, {Auger},
  {Barucci}, {Bertaux}, {Bertini}, {Besse}, {Bodewits}, {Capanna}, {Cremonese},
  {Da Deppo}, {Davidsson}, {Debei}, {De Cecco}, {Ferri}, {Fornasier}, {Fulle},
  {Gaskell}, {Giacomini}, {Groussin}, {Gutierrez-Marques}, {Guti{\'e}rrez},
  {G{\"u}ttler}, {Hoekzema}, {Hviid}, {Ip}, {Jorda}, {Knollenberg}, {Kovacs},
  {Kramm}, {K{\"u}hrt}, {K{\"u}ppers}, {La Forgia}, {Lara}, {Lazzarin},
  {Leyrat}, {Lopez Moreno}, {Magrin}, {Marchi}, {Marzari}, {Massironi},
  {Michalik}, {Moissl}, {Mottola}, {Naletto}, {Oklay}, {Pajola}, {Pertile},
  {Preusker}, {Sabau}, {Scholten}, {Snodgrass}, {Thomas}, {Tubiana}, {Vincent},
  {Wenzel}, {Zaccariotto}, \& {P{\"a}tzold}}]{Sierks:2015sb}
{Sierks}, H., {Barbieri}, C., {Lamy}, P.~L., {et~al.} 2015, Science, 347,
  aaa1044

\bibitem[{{Singer} {et~al.}(2016){Singer}, {McKinnon}, {Greenstreet},
  {Gladman}, {Parker}, {Robbins}, {Schenk}, {Stern}, {Bray}, {Spencer},
  {Weaver}, {Beyer}, {Young}, {Moore}, {Olkin}, {Ennico}, {Binzel}, {Grundy},
  \& {New Horizons Geology Geophysics and Imaging Science Theme
  Team}}]{Singer:2016sm}
{Singer}, K.~N., {McKinnon}, W.~B., {Greenstreet}, S., {et~al.} 2016, in
  AAS/Division for Planetary Sciences Meeting Abstracts, Vol.~48, AAS/Division
  for Planetary Sciences Meeting Abstracts

\bibitem[{{Singer} {et~al.}(2015){Singer}, {Schenk}, {Robbins}, {Bray},
  {McKinnon}, {Moore}, {Spencer}, {Stern}, {Grundy}, {Howett}, {Dalle Ore},
  {Beyer}, {Parker}, {Porter}, {Zangari}, {Young}, {Olkin}, \&
  {Ennico}}]{Singer:2015ss}
{Singer}, K.~N., {Schenk}, P.~M., {Robbins}, S.~J., {et~al.} 2015, in
  AAS/Division for Planetary Sciences Meeting Abstracts, Vol.~47, AAS/Division
  for Planetary Sciences Meeting Abstracts, 102.02

\bibitem[{Speith(2006)}]{Speith:2006}
Speith, R. 2006, Habilitation, University of T\"ubingen

\bibitem[{{Tsiganis} {et~al.}(2005){Tsiganis}, {Gomes}, {Morbidelli}, \&
  {Levison}}]{Tsiganis:2005tg}
{Tsiganis}, K., {Gomes}, R., {Morbidelli}, A., \& {Levison}, H.~F. 2005, \nat,
  435, 459

\bibitem[{Weissman {et~al.}(2004)Weissman, Asphaug, \&
  Lowry}]{Weissmann:2004wa}
Weissman, P.~R., Asphaug, E., \& Lowry, S.~C. 2004, Comets II, 337

\bibitem[{{Yamamoto}(1985)}]{Yamamoto:1985}
{Yamamoto}, T. 1985, \aap, 142, 31

\end{thebibliography}
%

\end{document}